\documentstyle[epsf]{mn}

\newcommand{\kms}{{\rm\ kms}$^{-1}$}
\newcommand{\oiii}{[O\,{\sc iii}] 5007-\AA}
\newcommand{\ngc}{NGC~4051}
\title[Evidence for an outflow from NGC 4051]{Evidence for an outflow from the Seyfert galaxy NGC~4051}
\author[]
       {P.E. Christopoulou,$^{1}$ A.J. Holloway,$^{2}$ W. Steffen,$^{2}$ C.G. Mundell,$^{3}$\cr A.H.C. Thean,$^{3}$ C.D. Goudis,$^{1}$ J. Meaburn,$^{2}$ and A. Pedlar$^{3}$\\
    $^{1}$Department of Physics, University of Patras, Patras 26500, Greece\\
    $^{2}$Department of Physics and Astronomy, University of Manchester, Schuster Laboratory, Oxford Road, Manchester M13 9PL, UK\\
    $^{3}$Nuffield Radio Astronomy Laboratories, University of Manchester, Jodrell Bank, Macclesfield, Cheshire SK11 9DL , UK\\  
}
\date{Accepted for publication August 1996}

\begin{document}

\maketitle

\begin{abstract}

New observations using narrow band imaging, long-slit spectroscopy and
MERLIN observations of the nuclear region of the Seyfert galaxy
NGC~4051 have been made.  An edge brightened, triangular region of
ionized gas extending 420 pc from the centre of the galaxy has been
detected. Long-slit spectra of this ionised gas, taken at 1.5\arcsec\
from the core, show the \oiii\ emission line to consist of two
velocity components, both blue-shifted from the systemic radial
velocity, with velocity widths of 140\kms\ and separated by
120\kms. This region is co-spatial with weak extended radio emission
and is suggestive of a centrally driven outflow.  The \oiii\ line
spectrum and image of this region have been modelled as an outflowing
conical structure at 50\degr\ to the line of sight with a half opening
angle of 23\degr .

 In addition to the extended structure, high resolution MERLIN
observations of the 18-cm nuclear radio emission reveal a compact
(1\arcsec) radio triple source in PA 73$^{\circ}$. This source is
coincident with the HST-imaged emission line structure. These high
resolution observations are consistent with a more compact origin of
activity (i.e. a Seyfert nucleus) than a starburst region.

\end{abstract}




\section{Introduction}

Anisotropic emission in Active Galactic Nuclei (AGN), and, in
particular, Seyfert galaxies, has become fairly well established
although the mechanisms of fuelling and collimating ejecta remain
problematic.  In Seyfert galaxies, the closest and a very common type
of AGN, both the radio and optical emission are often observed to be
emitted in a collimated fashion.

Seyfert galaxies are classified, according to the width of their
emission lines (Khachikian \& Weedman 1971), into two main types:
Type 1's (with broad permitted and narrow forbidden lines) and Type
2's (with narrow permitted and forbidden lines).  The discovery of a
hidden Broad Line Region (BLR) (Antonucci \& Miller 1985) and
non-stellar continuum in the polarised flux spectrum of the Seyfert
type 2 galaxy, NGC 1068, led to the present day Unified Schemes in
which the different observed properties of Seyfert types 1 and 2 may
be accounted for by viewing angle. Antonucci \& Miller (1985)
suggested that the nucleus in Seyfert type 2's are obscured from
direct view by an optically and geometrically thick disk or torus but
that the hidden BLR can be seen when nuclear radiation is scattered,
by electrons above and below the poles of the torus, into the
observer's line of sight. This obscuring torus would also give rise to
an anisotropic nuclear radiation field.

Direct optical evidence for this anisotropic radiation field was
provided by the discovery of the Extended Narrow Line Region (ENLR)
(Unger et al. 1987). The physical and kinematic properties of the
ENLR (e.g., FWHM $\le$ 45 km s$^{-1}$, \oiii/H$_{\beta}$ $\sim$ 10)
implied that it is ambient galactic gas photoionized by the nuclear
UV continuum radiation.  This was confirmed by the discovery of a
number of extended emission line regions (Meaburn, Whitehead \&
Pedlar 1989, Pogge 1989, Pedlar et al. 1989, Tadhunter \& Tsvetanov
1989, Unger et al. 1992, Wilson \& Tsvetanov 1994), ranging in size
from $\sim$70 pc to $\sim$20 kpc (Wilson \& Tsvetanov 1994), consistent
with ionisation by a UV radiation cone.

Until recently the relatively poor angular resolutions of ground based
optical measurements and small linear extents of the NLRs resulted in
limited evidence for {\em NLR} `cones' in Seyferts. However, the
discovery of NLR `cones' in NGC 4151 (Evans et al. 1993, Boksenberg et
al. 1995), NGC 3281 (Storchi-Bergman, Wilson \& Baldwin 1992), and
NGC 3227 (Mundell et al. 1995) suggests that this phenomenon may be
more widespread.

It should be noted that an ENLR is physically distinct from a Narrow
Line Region (NLR), which has broader emission lines (several hundred
\kms), is more directly affected by the active nucleus and may involve
different physical processes to the ENLR. In some Seyferts, the
optical emission may be a signature of the physical interaction
between the radio ejecta and the surrounding medium (e.g., Whittle et
al. 1986, Pedlar et al. 1989) as in the model proposed by Taylor,
Dyson \& Axon (1992).  Therefore, any detection of emission line
regions, consistent with a cone of ionising UV, should be considered
carefully together with kinematic information to determine its true
nature.

Although Seyferts are radio `quiet', they are not radio silent and
many Seyferts exhibit radio anisotropy in the form of highly collimated
radio jets (e.g., Wilson \& Ulvestad 1987, Pedlar et al. 1993,
Kukula et al. 1995). Although radio jets have typical opening angles
of only a few degrees, compared with ionisation cones of up to
$\sim$100 degrees, there are good alignments between ENLRs and radio jet
axes in a number of Seyferts (Wilson \& Tsvetanov 1994). The
alignments suggest that the radio ejecta and UV photons are collimated
by the same, or co-planar, structures with no significant relative
precession (Tsvetanov, Kriss \& Ford 1994). In those Seyferts which exhibit
misaligned radio jets and and ionisation wedges, it may that the UV
cone of radiation (which is assumed to share the same collimation
direction as the radio jet) points partly out of the galactic disk.
The UV cone then grazes the neutral galactic disk producing an ionised
wedge (in the disk) that is misaligned from the radio jet due to
projection effects on the sky (Pedlar et al. 1993, Robinson et
al. 1994).

Not all Seyferts have well-collimated radio jets and a distinction
should be made between those with jets and those with more diffuse,
extended radio emission. Baum et al. (1993) find large-scale, diffuse,
sometimes `bubble-like', radio emission from 12 Seyferts, including
NGC 4051. They note that the large-scale, `extra-nuclear' radio
emission is randomly oriented with respect to any small-scale nuclear
radio source axis and tends to align with the minor axis of the host
galaxy. They attribute this diffuse component of the radio emission to
radio-emitting superwind that has been produced and swept out along
the galaxy's minor axis by a circum-nuclear starburst.

\ngc\ has been the subject of a number of studies in the past and in 
Table 1 we list the morphological parameters from previous
observations.  At radio wavelengths structure has been seen on a
number of scales. The VLA observations by Ulvestad \& Wilson (1984) at
6cm detect the 0.4\arcsec\ double source at PA 78\degr$\pm$6\degr,
with more extended emission ($\sim$2\arcsec) to the southwest in their
20cm map. On a much larger scale, Baum et al. (1993) detect 3 regions
of 6cm radio emission: a 'banana-shaped' nuclear region (15\arcsec\ in
extent) along PA 32\degr (the minor axis of the galaxy), an
extra-nuclear diffuse radio emission region from the spiral arms and
large-scale radio emission from the entire galaxy disk. The 8.4 GHz
VLA C-array maps from Kukula et al. (1995) shows a bright elongated
source at PA 37\degr\ with a total extent of 15\arcsec.

The \oiii\ image of Haniff, Wilson \& Ward (1988) contains a compact
emission line region of overall extent 3\arcsec\ at PA
81\degr$\pm$5\degr, which suggests that the orientation and extent of
the high excitation emission line gas are essentially the same as
those of the nuclear radio continuum emission. The main characteristic
of the narrow line profiles of \ngc\, present in all the forbidden
lines, is the strong blue wing extending to velocities of -800\kms
with substructures especially evident in the profiles of [O\,{\sc
iii}] 4959,5007-\AA (Veilleux 1991).

\begin{table*}
{\centering
\caption{Morphological parameters for NGC 4051}
\label{params}
\begin{tabular}{lll}
Other name & UGC 07030 & (Uppsala General Catalogue of Galaxies - UGC, Nilson 1973)\\
Seyfert type & Type 1 & (Adams 1977)\\
 & Type 1.5 & (Dahari \& De Robertis 1988)\\
Host galaxy & SB & (Catalogue of Principal Galaxies - CPG, Paturel et al. 1989)\\
 & Sb or SBc & (UGC)\\
Magnitude (V) & 12.9 & (Veron-Cetty \& Veron 1985)\\
Inclination & -40\degr & (Adams 1977)\\
Optical major axis PA & 135\degr & (CPG, UGC)\\
Optical minor axis PA & 32\degr & (CPG, UGC)\\
Heliocentric redshift & 0.0023 & (Dahari \& De Robertis 1988)\\
Vr$^{a}$ (\kms) & 613 & (Kukula et al. 1995)\\
Vsys (\kms) & 726 & (Ulvestad \& Wilson 1984)\\
Optical flux density (mJy) & 200 & (Kukula et al. 1995)\\
F(\oiii) ($10^{-14}$ergs s$^{-1}$ cm$^{-2}$) & 52.1 & (Dahari \& De Robertis 1988)\\
\end{tabular}
}

\medskip

$^{a}$ Heliocentric recession velocity.
\end{table*}

In the present paper we describe new optical and radio observations of
the Seyfert galaxy \ngc. We adopt a distance of 9.7 Mpc, using a value
of H$_{0}$=75\kms Mpc$^{-1}$ and its redshift of 726\kms (Ulvestad \&
Wilson 1984). One arcsecond therefore equals 47 pc at the galaxy.

\section{Observations and results}

\subsection{MES \oiii\ narrow band imaging}

Narrow band imaging in the light of the \oiii\ emission line was
carried out on 1993 June 11 using the Manchester Echelle Spectrograph
(MES - Meaburn et al. 1984) in its imaging mode, combined with the
f/15 Cassegrain focus of the 2.5m Isaac Newton Telescope. An EEV CCD
(1280x1180 pixels) detector was used to provide a binned 2x2 array of
640x590 pixels which, at the f/15 focus, provided a pixel scale of
0.255\arcsec/pixel. During these observations the seeing was typically
0.8\arcsec.

In the imaging mode a plane mirror is inserted to exclude the echelle
grating and a clear aperture replaces the entrance slit. Four
exposures, totalling 7600s, were made of the galaxy using a filter
with a central wavelength of 5050\AA\ and a bandwidth of 70\AA\ to
include the redshifted \oiii\ line. This filter also included the
[O\,{\sc iii}] 4959-\AA\ line but only at a level of $\sim$8 per cent
of the total flux. To enable the continuum light to be subtracted from
this image two images were taken using a V band (1000\AA\ bandwidth)
filter with a total exposure time of 300s.

The reduction of the data was carried out at the University of
Manchester node of the UK STARLINK computer network using the {\sc
figaro, ccdpack} and {\sc kappa} packages. The images were processed
in the usual way and then images through each filter were co-added to
improve the signal-to-noise ratio. The images were aligned using the
compact peak in emission from the nucleus of the galaxy. The \oiii\
image was then obtained by the subtraction of the V-band image from
the \oiii\ + continuum image, being scaled, in the absence of any
stars, so that the galactic disk disappeared in the final image.

The V-band image of the galaxy and continuum subtracted \oiii\
emission line image are displayed using a logarithmic scale and are
shown in Fig. 1.

\begin{figure*}
\centering
\mbox{\epsfclipon\epsfxsize=7in\epsfbox[20 74 500 326]{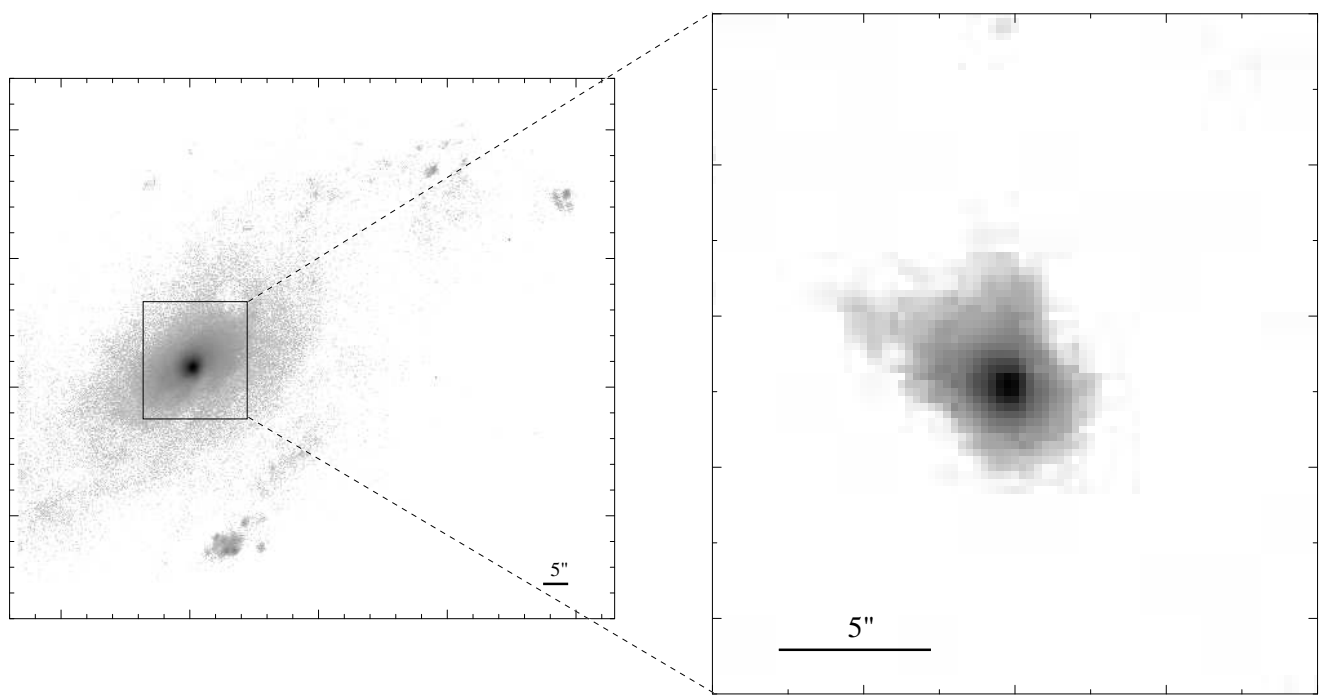}}
\caption{MES INT V-band image (left) and \oiii\ image after continuum subtraction (right) of NGC 4051, displayed in negative using a logarithmic scale}
\label{vband}
\end{figure*}

\subsection{MES spectrometry}

Spatially resolved spectral observations were made on 1993 June 11
using the same instrumental setup as described above. A single
longslit was used with the 5050\AA\ central wavelength filter and the
redshifted \oiii\ line observed in the 113th echelle order. The
integration time was 1800s, using the EEV CCD binned this time twice
in the spatial direction and by three times in the spectral
dimension. This results in an array of 640 pixels along the slit
length with a scale of 0.255\arcsec/pixel and 393 pixels in the
wavelength axis with a spectral resolution of $\sim$ 9\kms\. The slit
of the spectrometer was rotated to a position angle of 27.5\degr\ to
align it with that of the radio emission and centered on the nucleus
of the galaxy. A slit width of 1.62\arcsec\ on the sky was used and
the seeing was typically 0.8\arcsec.

The data were reduced using the STARLINK {\sc figaro, ccdpack, kappa}
and {\sc twodspec} packages. A tungsten lamp was used to flat-field
the spectrum and a Th-Ar arc lamp was used to calibrate the wavelength
to $\pm$1\kms\ accuracy. The rest wavelength for the \oiii\ line was
taken as 5006.85\AA.

\begin{figure*}
\centering
\mbox{\epsfclipon\epsfxsize=7in\epsfbox[53 100 413 368]{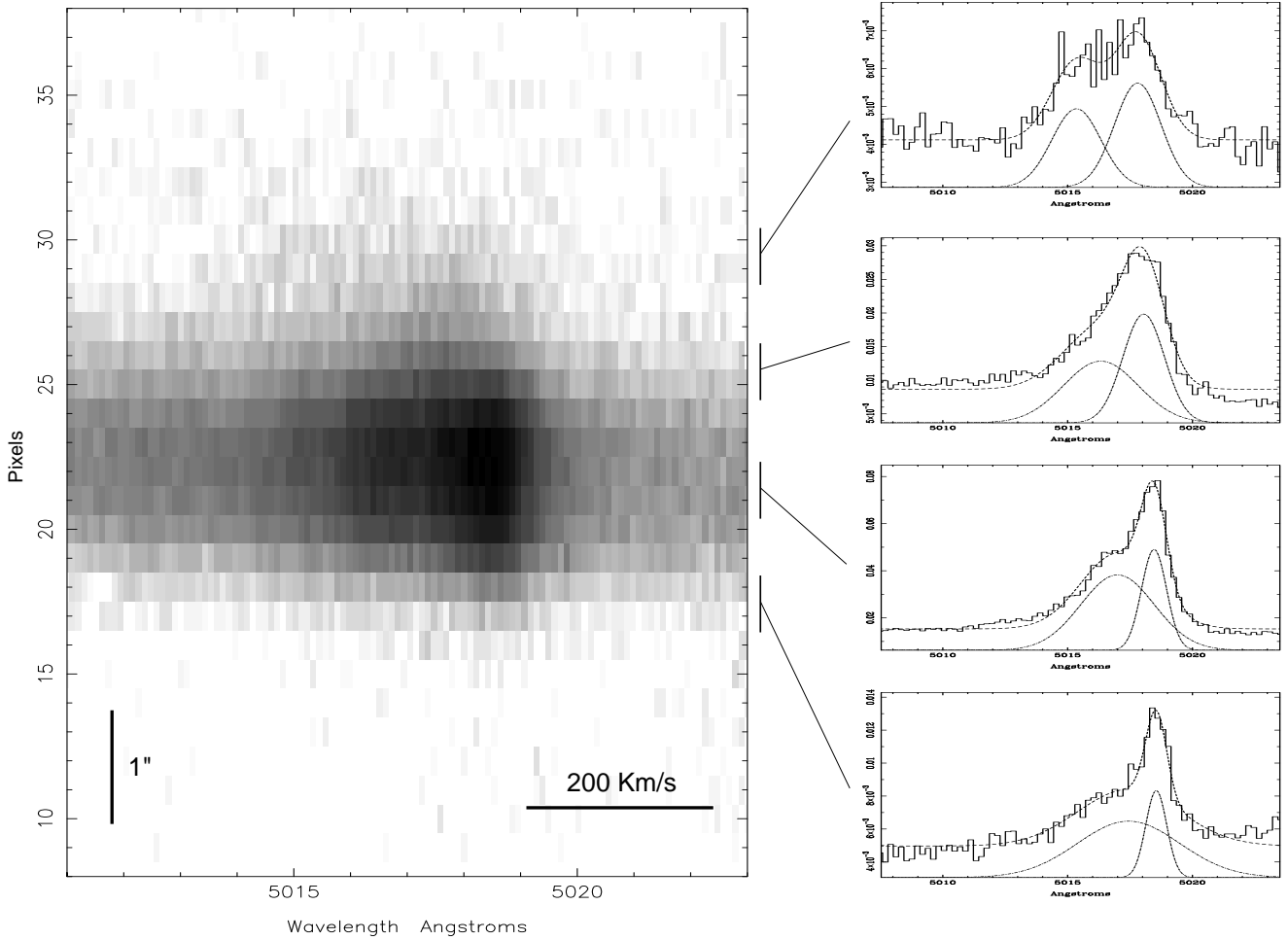}}
\caption{INT MES position-velocity map of \oiii\ emission along PA 27.5\degr\ displayed on a logarithmic scale. Spectra averaged over 0.51\arcsec\ regions separated by 1\arcsec\ are shown on the right with two-component Gaussian fits. North is to the top of the map. }
\label{oiiispec}
\end{figure*}

The resulting position-velocity (pv) array is shown in
Fig. \ref{oiiispec} with spectra averaged over 0.51\arcsec\ regions
along the slit and with fits from a two component Gaussian model.

\subsection{Radio emission in NGC 4051 - 18cm MERLIN maps}

\begin{figure}
\centering
\mbox{\epsfclipon\epsfxsize=3.3in\epsfbox[40 121 570 687]{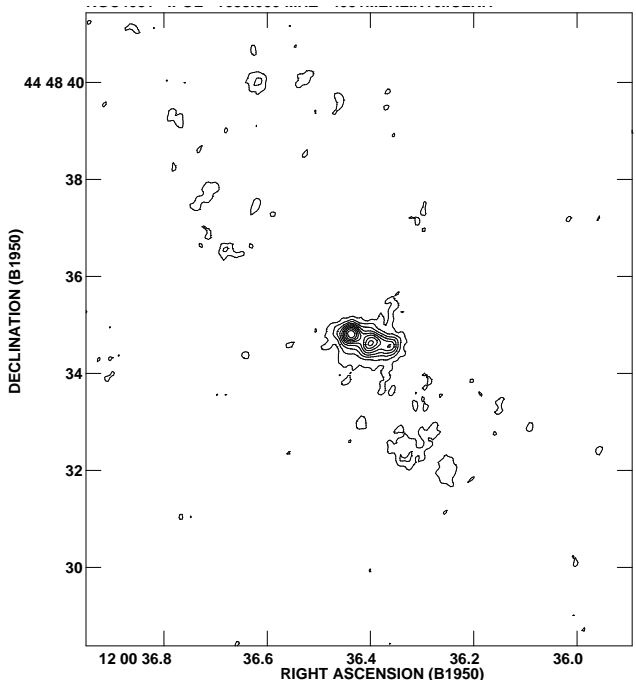}}
\caption{18-cm MERLIN naturally weighted map of NGC 4051 with contours at 14.52x( 1, 2, 3, 4, 5, 6, 7, 8, 9) mJy}
\label{merlin}
\end{figure}

NGC 4051 was observed by MERLIN in October 1993 at 1658MHz. An 18 hour
observing run was carried out in phase referencing mode using the
nearby calibrator 1200+468. The flux density was determined from
observations of 3C286. The data were imaged using natural weighting
(Fig. \ref{merlin}) resulting in an angular resolution of 0.35\arcsec
x 0.31\arcsec. The MERLIN radio maps show a 0.8\arcsec\ triple source
extended along PA 73\degr. The flux densities of the radio components
are given in Table 2.  In the naturally weighted image there is weak
emission extending over 8\arcsec\ in PA $\sim 30\degr$, consistent
with lower resolution results reported by Baum et al. (1993) and
Kukula et al. (1995).

\subsection{Archive HST data}

\begin{figure}
\centering
\mbox{\epsfclipon\epsfxsize=3.3in\epsfbox[52 93 347 303]{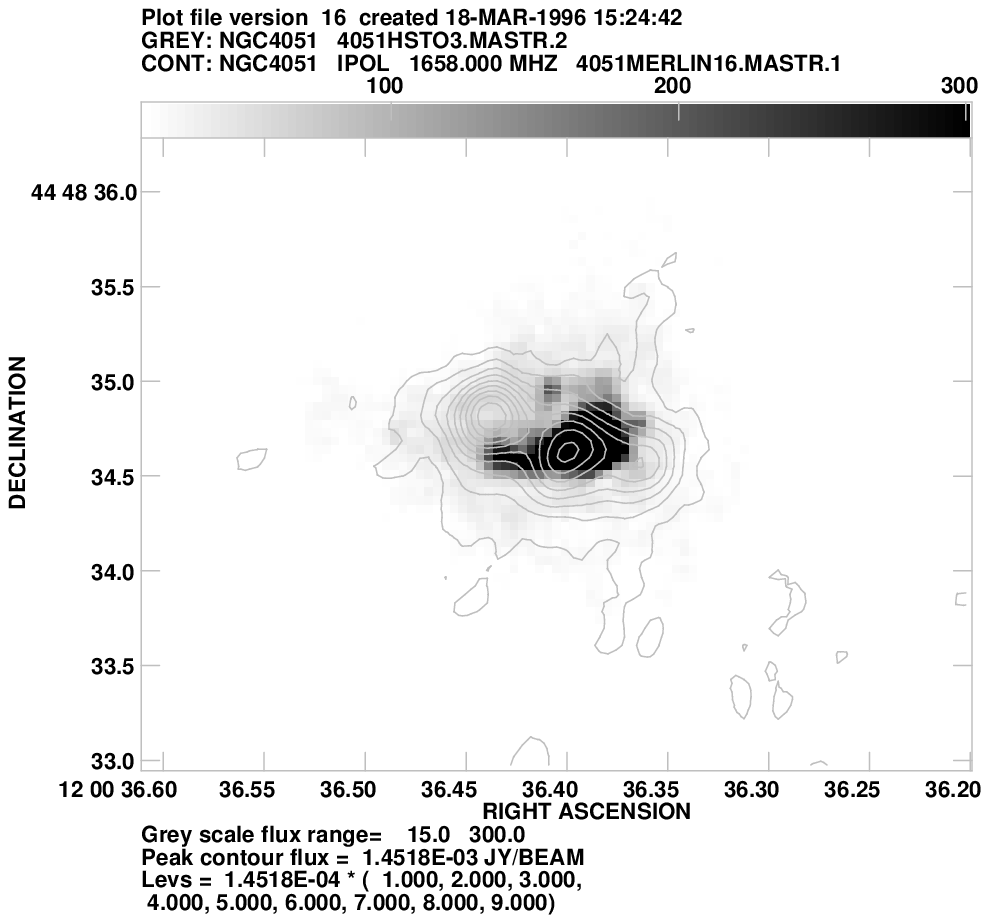}}
\caption{The HST F502N image is plotted as a greyscale with the MERLIN 18-cm map (Fig. 3) overlaid as contours.}
\label{hst}
\end{figure}

For comparison with the MERLIN observations, Hubble Space Telescope
(HST) images were obtained from the HST data archive. The observation
was carried out using the F502N filter (including \oiii\ in the
bandpass) with the Wide Field Planetary Camera I on 20th June 1991
(proposal no. 1036, H.C. Ford).  This image has then been deconvolved
using a model point spread function generated with the program {\sc
Tiny Tim}, and applying the Lucy deconvolution routine within the IRAF
package {\sc stsdas}. The image is shown in Fig. \ref{hst} with the
MERLIN 1658 Mhz map overlaid as contours. A central 0.6\arcsec\ linear
feature is visible and more diffuse emission is seen out to a distance
of 3\arcsec\ in the HST image.

\section{Discussion}

\subsection{The triangular emission line region}

\begin{figure}
\centering
\mbox{\epsfclipon\epsfxsize=3.3in\epsfbox[56 154 541 609]{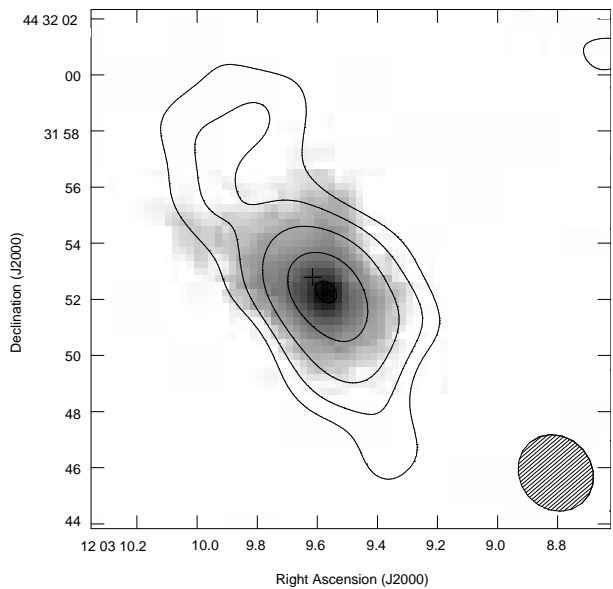}}
\caption{MES \oiii\ image (Fig. 1) overlaid with contours of the 8 Ghz C-array VLA map of Kukula et al. (1995)}
\label{marek}
\end{figure}

The \oiii\ MES image of \ngc\ is shown in Fig. \ref{marek} with the 8 GHz VLA
C-array map of Kukula et al. (1995) overlaid.

From the \oiii\ imaging (Fig. 1) we can clearly see the presence of a
9\arcsec(420 kpc) ionised wedge in the centre. The wedge is bisected
at a position angle of 33\degr, similar to that of the large scale
radio axis (Fig. \ref{marek}). The wedge is also therefore pointing
along the minor axis of the galaxy (to within 1$\degr$) as defined by
the V-band image. The linear extent is similar to the radio emission
seen in this direction, and the sharp edges of the \oiii\ wedge seem
to follow the outline of the radio emission. The wedge is edge
brightened and if its apex is assumed to be coincident with the
nucleus of \ngc\ then it has an apparent opening angle of 55\degr. No
extended emission is detected to the southwest of the nucleus.

In the nuclear region of \ngc\ (central 0.5\arcsec) the width (FWHM)
of the \oiii\ line is 240\kms, typical of NLR gas, though from the
Gaussian fitting it does appear to have both a narrow and broad
component.  Spectra taken 1.5\arcsec\ to the northeast of the nucleus
(top plot, Fig. 2) show evidence of two equal components separated by
$\sim$ 120\kms, where both lines are $\sim$140\kms\ wide and are
blue-shifted by 80 and 225\kms\ from the systemic radial velocity.

Similar linewidths and line-splitting are seen in the ionised wedges
in NGC 3281 (Storchi-Bergman et al. 1992) and NGC 3227 (Mundell et
al. 1995, Mundell 1995).  Storchi-Bergman et al. (1992) propose a
model in which outflow is taking place along an axis at a large angle
to the plane of the galaxy. Unlike NGC 4051 and NGC 3227, in which
both line components are blue-shifted, one of the components in NGC
3281 is red-shifted and the other blue-shifted. Therefore, they
propose two possible geometries to explain their observed velocities;
one where the cone axis is perpendicular to the plane of the galaxy
and the other where the cone axis makes an angle of 45\degr\ with the
disk.  In their model, gas flows out within a conical envelope or on
the surface of a hollow cone. The line-splitting is then attributed to
the two components of velocity, along the line of sight, from the near
and far sides of the cone.

We will now apply a similar model to \ngc\ but, since both components
are blue-shifted, the orientation of the outflow cone may be
determined unambiguously.

\subsection{Kinematic modelling of the extended emission line region}

\begin{figure}
\centering
\mbox{\epsfxsize=3.3in\epsfbox[49 111 386 404]{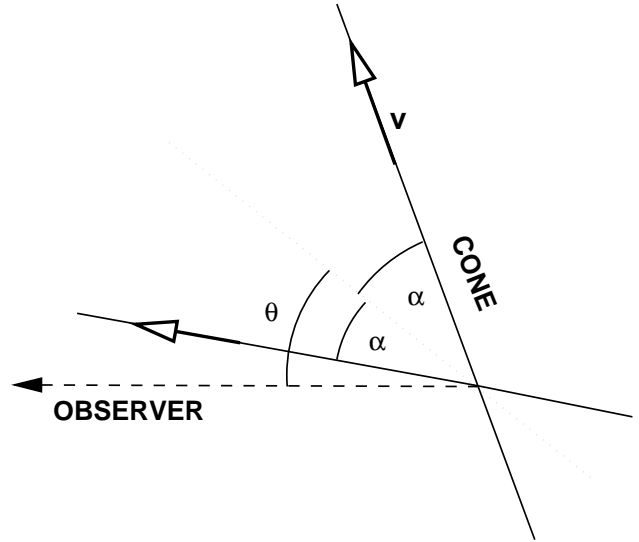}}
\caption{A schematic diagram illustrating the cone geometry.}
\label{schmod.fig}
\end{figure}

Our spectra reveal two velocity components, both blue shifted, which
suggests that the axis of the emission cone is inclined towards the
observer as shown in Fig. \ref{schmod.fig}. Here the velocity along
the sides of the cone is defined as $v$, with the inclination of the
axis of the cone at $\theta$ degrees to the line of sight and the cone
having a half opening angle (for simplicity, hereafter `opening
angle') of $\alpha$ degrees.

Assuming the cone to be circular in cross-section, the true opening
angle of the cone, $\alpha$, is

\begin{equation}
\tan\alpha \approx \tan\alpha_{obs}\sin\theta,
\end{equation}
\noindent
where $\alpha_{obs}$ is the observed opening angle of the cone.

If we assume the double velocity components present in the spectrum to
be coming from opposite sides of the cone and the true velocity
direction is along the sides of the cone, then the observed velocities
$v_{1}$ and $v_{2}$ are defined by,

\begin{equation}
v_{1}=v\cos\left(\theta-\alpha\right),\\
v_{2}=v\cos\left(\theta+\alpha\right).
\end{equation}

Using our values for the observed velocities of the Gaussian
components in the spectrum $v_{1}$=225\kms, $v_{2}$=80\kms and
$\alpha_{obs}$=30\degr\ we can solve equations (1) and (2) to obtain
$\theta=48\degr$, $\alpha=23\degr$ and a velocity along the cone of
$v$=245\kms (see Fig. \ref{schmod.fig}). We use the \oiii\ line
profiles as measured at more than 1.5\arcsec\ from the nucleus in
order to minimize contamination from the core region.

\begin{figure}
\centering
\mbox{\epsfclipon\epsfxsize=3.3in\epsfbox[22 211 563 513]{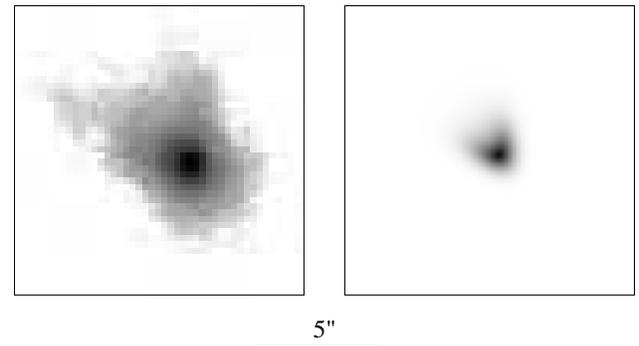}}
\caption{A comparison of the observed MES \oiii\ emission line wedge (left) with the image generated from the model (right).}
\label{modim}
\end{figure}

\begin{figure}
\centering
\mbox{\epsfxsize=3.3in\epsfbox[130 95 434 227]{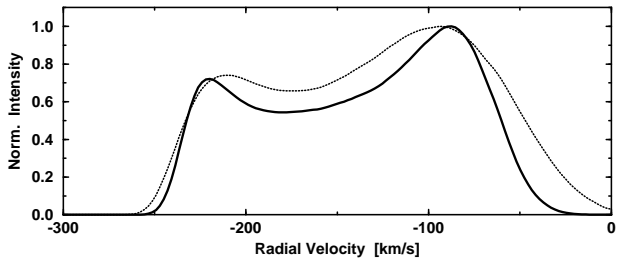}}
\caption{The model spectrum at 1.5\arcsec\ from the nucleus. The dotted line is for a wall thickness comparable to the local cone radius, whereas the solid line corresponds to a wall thickness of half of the local cone radius. Compare this with the spectral plot at the top of Fig. 2.}
\label{modspec}
\end{figure}

For simplicity, we used a Gaussian emissivity distribution for the
wall of the cone (i.e. the emissivity drops smoothly away from the
local radius of the wall).  We assume that the emitting gas moves
radially outwards from the vertex of the cone. We also assume that the
emissivity along the jet axis drops with a Gaussian distribution (FWHM
= 127~pc).  Using the parameters derived from the spectrum and ground
based image, this model yields a reasonably good approximation to the
data (Figs. \ref{modim} and \ref{modspec}). However, the apparent
opening angle in the simulated image and the exact shape of the
spectrum both depend on the thickness of the wall of the cone.

Similarly, the width of the spectral line coming from each side of the
cone is determined by the width of the wall. Note that in this model
the side of the cone which is best aligned with the line of sight will
have a smaller spectral width than the other. This effect can be seen
in Fig. \ref{modspec}, where spectra are drawn with different wall
thicknesses. Here a slit of width 2.5\arcsec\ was set to be aligned
with the cone axis, and a section between 1.5\arcsec\ and 2.1\arcsec\
from the core was used to produce this spectrum. The spectral
resolution was 9\kms\ and the spatial resolution was 0.8\arcsec. A
larger model slit width was used than in the observations to
compensate for the image motion due to seeing, which was not taken
into account explicitly. The solid line in Fig. \ref{modspec} is for a
FWHM equal to half the local radius of the cone, whereas the dashed
line is for a FWHM twice as large. Because of the radial motion of the
gas, thicker walls result in wider lines. This simulation can be
compared with the spectrum of Fig. \ref{oiiispec} taken 1.5\arcsec\
from the core (top spectral plot). In Fig. \ref{modim} a model image
of a wedge is shown with the observed MES \oiii\ image for
comparison. The agreement is very good, considering the simplicity of
this kinematic model. The main differences arise from not taking into
account the central point source which produces a bright core at the
vertex of the cone and a bright feature in the longslit
spectrum. Hence, spectra produced from cuts through the model cone
considerably closer to the core would not be expected to agree with
the observed spectra. However, at 1.5\arcsec (roughly 2 widths of the
point spread function) we can expect a very reduced influence of the
core emission, which affects mainly the height ratio of the spectral
cone components. In our model, the height ratio is determined mainly
by the emissivity variation along the axis of the cone.

Varying the parameters like the outflow velocity, the FWHM of the
emissivity along the axis, the opening angle or the angle to the line
of sight will produce results consistent with the observations with
ranges of approximately plus or minus $\Delta v$=30~\kms, $\Delta{\rm
FWHM}_z$=20~pc, $\Delta\alpha=2\deg$, and $\Delta\theta=5\deg$,
respectively.  Based on this model, we suggest that in NGC~4051 the
\oiii\ wedge represents a radial conical outflow with a half opening
angle near 23\degr\ and a velocity of approximately 250\kms\ at an
angle to the line of sight of roughly 50\degr.

Note that in the two-component fit to the spectrum in
Fig. \ref{oiiispec}, the component which is blue-shifted most gradually
moves towards the systemic radial velocity and broadens as the core of the
galaxy is approached. This does, of course, not happen in our model
and may be due to a significant contribution of the redder line to the
Gaussian fit of the bluer one. Unfortunately, the data quality does
not allow a well constrained three component fit to separate the core
component from the cone components.

An alternative reason for this line shift may be that the cone does
not have straight sides and may even change direction closer to the
centre of the galaxy. This could also be combined with inhomogeneities
in the cone wall. An observational indication of this is the position
angle of the inner radio source, which is noticeably different from
the outer one and the structure of the emission line gas in the
HST-image (compare Fig. \ref{vband} and \ref{hst}). However, these
features are still consistent with the overall picture of an edge
brightened, roughly conical outflow with a velocity around or in
excess of 200\kms.

Our conical model should be compared with an alternative physical
model which is possible at least in principle.  In this alternative
model the observed wedge would actually be the inner edge of a plasma
cloud which is expanding perpendicularly to its surface. A similar
analysis to the one presented above, shows, however, that this is not
consistent with the blue-shifted double-line, except if the `cone' is
on the far side of the galaxy. In this case, the observer's line of
sight would have to point inside the cone. Assuming a roughly uniform
azimuthal emission distribution, this is not consistent with the
observed edge brightening of the \oiii\ wedge. We therefore favour the
scenario of a conical outflow with a vertex near the centre of the
galaxy. Our simple outflow model does not provide a physical reason
for the large scale ejection of the optically emitting gas. However,
we speculate that it might be related to the collimated radio outflow or
a large scale `superwind' (Baum et al. 1993).

A trend seems to be emerging that an increasing number of cones
detected in Seyfert galaxies are consistent with NLR rather than ENLR
gas. The kinematics of the gas in these cones are more indicative of
outflow rather than of anisotropic photoionisation of ambient galactic
gas. The gas may be partially photoionized by the active nucleus but,
if the radio ejecta are interacting with the optical gas, shock
ionisation may also be present. The structure and kinematics, in
particular line-splitting, of the \oiii\ emitting gas in the wedges in
NGC 4051, NGC 3281, and NGC 3227 is consistent with conical outflow along
the galactic density gradient.

\subsection{The compact radio structure}

Baum et al. (1993) have suggested that the extended radio emission
seen in NGC 4051 is a consequence of outflow caused by a starburst
driven `superwind' which causes the optically emitting gas to flow out
also. The ratio of 60 $\mu$m to radio flux (Condon, Anderson \&
Broderick 1995) is also consistent with a starburst rather than a
compact active nucleus. In general, starbursts are associated with
diffuse radio emission, whereas Seyferts show collimated structures.

\begin{table}
{\centering
\caption{Radio parameters for NGC 4051}
\label{specind}
\begin{tabular}{cccc}
Component & Flux (mJy) & Flux (mJy)  & Spectral Index\\
 & 1.658 GHz & 8.439 GHz & (1.658/8.439)\\
East & 1.6 & 0.05 & 2.1\\
Central & 1.6 & 0.50 & 0.7\\
West & 0.7 & 0.05 & 1.6\\
\end{tabular}
}
\end{table}

However, the high resolution maps made using MERLIN at 1658 Mhz
(Fig. \ref{merlin} reveal a compact (0.8\arcsec, 38 pc) triple source
which favours collimated ejection in PA $73\degr$. The 8GHz VLA A
array image (Kukula et al. 1995) also shows weak structure consistent
with the above triple, and by comparison with the 1658 Mhz image we
can deduce the spectral indices of the components (see Table
\ref{specind}). It is clear that the central component has the
flattest spectral index of the triple, which would tentatively
identify it as the AGN. The newly detected third component could
therefore be part of a counter jet and the extended radio emission
could represent a continuation of the central collimated flows
somewhat analogous to Fanaroff-Riley 1 radio galaxy structures.
However, in Seyfert galaxies the extended radio emission could be due
to a debris of relativistic particles from frustrated jets flowing
down the galactic density gradient similar to one of the models
proposed for Mkn6 (Kukula et al. 1996). The fact that the HST
structure (Fig. \ref{hst}) bends towards PA $\sim 30\degr$ may also be
evidence in favour of such a process.

The HST image (Fig. 4) shows a strong emission line structure of
similar size to the radio triple with the western end of the optical
feature being the brightest. Whilst the alignment of the HST image is
subject to a pointing error of $\sim$0.6\arcsec (Cox - private
communication), this structure probably represents the base of
the larger cone found in our ground based observations which engulfs
the more extended weak radio emission.  This is very similar to what
has been found in Mkn 6 (Kukula et al. 1996).

\section{Conclusions}

We have found an edge-brightened extended emission line region
exhibiting \oiii\ line splitting of up to $\sim$120\kms\ at
1.5\arcsec\ from the core.  We suggest that this is due to a conical
outflow with a flow velocity of approximately 250\kms\ with its axis
at roughly 50\degr\ to the line of sight. This engulfs the faint radio
emission found on the same scale.

The MERLIN results show evidence for a subarcsecond radio triple
source consistent with collimated ejection in PA 73\degr\ on a scale
less than 40 pc. The three components are misaligned with respect to
the large scale radio structure and optical emission line wedge,
though there is evidence that the structures on these scales are
linked.

Emission in the HST image seems to be the base of the emission line
wedge found in our ground based image. Although the sub-structure of
the HST and MERLIN images do not match each other, the overall
extension of the brightest emission is similar in the emission line
and the radio images and the compact radio emission again lies within
the area covered by optical emission.

Our present observations cannot distinguish between whether or not the
outflow is simply a continuation of the central collimated ejection
associated with the Seyfert nucleus or possibly due to a starburst on
a similar scale.  Higher sensitivity, radio and HST imaging, together
with dynamical studies are required to distinguish these
possibilities.

\section*{Acknowledgements}

We thank M. Kukula for the postscript map of the 8 Ghz VLA C-array
data.  We acknowledge the support of the staff at the INT during these
observations.  AJH, WS, CGM, and AHCT acknowledge the receipt of a
PPARC research studentship, associateship, fellowship and studentship
respectively. PEC and CDG acknowledge support by the British Council
during this work. We thank the anonymous referee for constructive
comments.

\end{document}